\documentclass[letterpaper]{article} 
\usepackage{aaai2026}  
\usepackage{times}  
\usepackage{helvet}  
\usepackage{courier}  
\usepackage[hyphens]{url}  
\usepackage{graphicx} 
\usepackage{natbib}  
\usepackage{caption} 
\usepackage{algorithm}
\usepackage{algorithmic}

\usepackage{booktabs}
\usepackage{multirow}
\usepackage{xcolor}     
\usepackage{colortbl}
\usepackage{makecell}
\usepackage{amsmath,amsthm,amssymb,amsfonts}

\theoremstyle{plain}
\newtheorem{definition}{Definition}

\usepackage[toc,page]{appendix} 
\renewcommand{\appendixpagename}{Appendix}

\usepackage{newfloat}
\usepackage{listings}
\DeclareCaptionStyle{ruled}{labelfont=normalfont,labelsep=colon,strut=off} 
\floatstyle{ruled}
\newfloat{listing}{tb}{lst}{}
\floatname{listing}{Listing}
\title{Differentiable Semantic Meta-Learning Framework for Long-Tail Motion Forecasting in Autonomous Driving}
\author{
    Bin Rao\textsuperscript{\rm 1},
    Chengyue Wang\textsuperscript{\rm 1},
    Haicheng Liao\textsuperscript{\rm 1},
    Qianfang Wang\textsuperscript{\rm 2},
    Yanchen Guan\textsuperscript{\rm 1},
    Jiaxun Zhang\textsuperscript{\rm 1},
    Xingcheng Liu\textsuperscript{\rm 1},
    Meixin Zhu\textsuperscript{\rm 3},
    Kanye Ye Wang\textsuperscript{\rm 4},
    Zhenning Li\textsuperscript{\rm 4} \thanks{Corresponding author: zhenningli@um.edu.mo}
}
\affiliations{
    \textsuperscript{\rm 1}State Key Laboratory of Internet of Things for Smart City, University of Macau, Macau SAR, China\\
    \textsuperscript{\rm 2}School of Civil Engineering and Transportation, South China University of Technology, Guangzhou, China\\
    \textsuperscript{\rm 3}School of Transportation, Southeast University, Nanjing, China\\
    \textsuperscript{\rm 4}State Key Laboratory of Internet of Things for Smart City and Departments of Civil and Environmental Engineering and Computer and Information Science, University of Macau, Macau SAR, China\\
    binrao.1@connect.um.edu.mo, yc47938@um.edu.mo, yc27979@um.edu.mo, qianfangwang@163.com, yc37976@um.edu.mo, yc47415@um.edu.mo, yc57431@um.edu.mo, meixin@seu.edu.cn, wangye@um.edu.mo, zhenningli@um.edu.mo
}

\begin{document}

\maketitle

\begin{abstract}
Long-tail motion forecasting is a core challenge for autonomous driving, where rare yet safety-critical events—such as abrupt maneuvers and dense multi-agent interactions—dominate real-world risk. Existing approaches struggle in these scenarios because they rely on either non-interpretable clustering or model-dependent error heuristics, providing neither a differentiable notion of “tailness” nor a mechanism for rapid adaptation. We propose SAML, a Semantic-Aware Meta-Learning framework that introduces the first differentiable definition of tailness for motion forecasting. SAML quantifies motion rarity via semantically meaningful intrinsic (kinematic, geometric, temporal) and interactive (local and global risk) properties, which are fused by a Bayesian Tail Perceiver into a continuous, uncertainty-aware Tail Index. This Tail Index drives a meta-memory adaptation module that couples a dynamic prototype memory with an MAML-based cognitive set mechanism, enabling fast adaptation to rare or evolving patterns. Experiments on nuScenes, NGSIM, and HighD show that SAML achieves state-of-the-art overall accuracy and substantial gains on top 1–5\% worst-case events, while maintaining high efficiency. Our findings highlight semantic meta-learning as a pathway toward robust and safety-critical motion forecasting.
\end{abstract}


\section{Introduction}
\begin{figure}[t]
  \centering
  \includegraphics[width=1\linewidth]{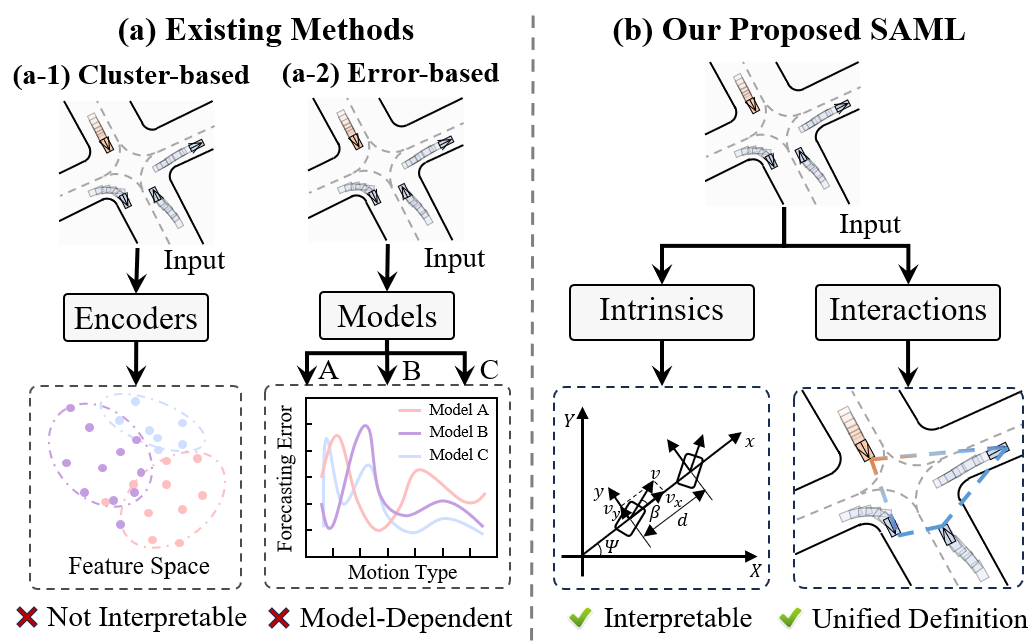}
   \caption{Conceptual comparison of (a) existing methods and (b) the proposed SAML framework. Existing methods detect long-tail events indirectly via non-interpretable clustering (a-1) or model-dependent error signals (a-2). In contrast, SAML (b) offers a principled, interpretable framework that quantifies a motion’s tailness from its intrinsic (dynamics, geometry, temporality) and interactive (local and global risk) properties, enabling robust long-tail forecasting.}
   \vspace{-9pt}
   \label{fig1}
\end{figure}

Motion forecasting is a foundational component of autonomous driving pipelines, yet current systems remain fragile in low-frequency, high-risk \textbf{long-tail} scenarios \cite{li2023graph-risk, wang2023fend}. These rare events—characterized by abrupt kinematic changes, complex multi-agent interactions, and atypical scene structures—are precisely the ones that most threaten safety and public trust. Despite steady progress on common cases, state-of-the-art predictors still degrade disproportionately on the tail, where data scarcity, semantic heterogeneity, and distributional shift collide \cite{salzmann2020trajectron++}.

A key reason for this brittleness is foundational: the community lacks a \textbf{principled}, \textbf{differentiable}, and \textbf{semantically} grounded definition of the “long tail”. Existing practices typically (i) cluster motions with non-interpretable heuristics \cite{makansi2021on-exposing, wang2023fend} or (ii) backsolve “hard cases” from model-specific forecasting errors \cite{zhang2024tract}. Both strategies are problematic. Clustering-based labels are hard to interpret and highly sensitive to hyperparameter choices, while error-based labels inherit the biases of a specific model, offering neither a generalizable notion of tailness nor a label that can guide end-to-end training (Figure \ref{fig1}). Crucially, both yield discrete, non-differentiable outputs, hindering gradient-based learning targeted at the tail.

To address these issues, we propose \textbf{SAML}, a \textbf{S}emantic-\textbf{A}ware \textbf{M}eta-\textbf{L}earning framework that redefines the long tail in a differentiable manner and learns to adapt to it. SAML rests on two pillars:

1) \textbf{Differentiable semantic definition of tailness}. We quantify a motion’s “tailness” through a principled set of intrinsic (kinematic dynamism, geometric complexity, temporal irregularity) and interactive (local interaction risk, global scene risk) metrics—each fully differentiable. A Bayesian Tail Perceiver aggregates these metrics into a Tail Index that is both uncertainty-aware and continuously optimizable.

2) \textbf{Tail-index–guided meta-adaptation}. The Tail Index steers a Meta-Memory Adaptation module that couples a dynamic prototype memory bank with an MAML-driven cognitive set mechanism. This design enables rapid, few-shot adaptation to emerging or sparsely observed long-tail patterns while mitigating bias toward majority behaviors.

The main contributions of this paper are threefold:
\begin{itemize}
    \item We operationalize tailness via semantically meaningful intrinsic and interactive metrics, aggregated by a Bayesian perceiver into a continuous Tail Index that unlocks end-to-end optimization for rare events.

    \item We introduce a Tail-Index–guided Meta-Memory Adaptation module that integrates a dynamic memory bank with a cognitive set mechanism inside a MAML framework, enabling fast adaptation to novel tail patterns.
    
    \item Extensive experiments on nuScenes, NGSIM, and HighD show that SAML achieves SOTA overall accuracy, with substantial gains in long-term horizons and worst-case (top 1–5\%) subsets, alongside competitive inference efficiency. Comprehensive ablations verify the necessity of each component, with the cognitive set mechanism yielding the largest impact on worst-case robustness.

\end{itemize}

\begin{figure*}[t]
  \centering
  \includegraphics[width=1\linewidth]{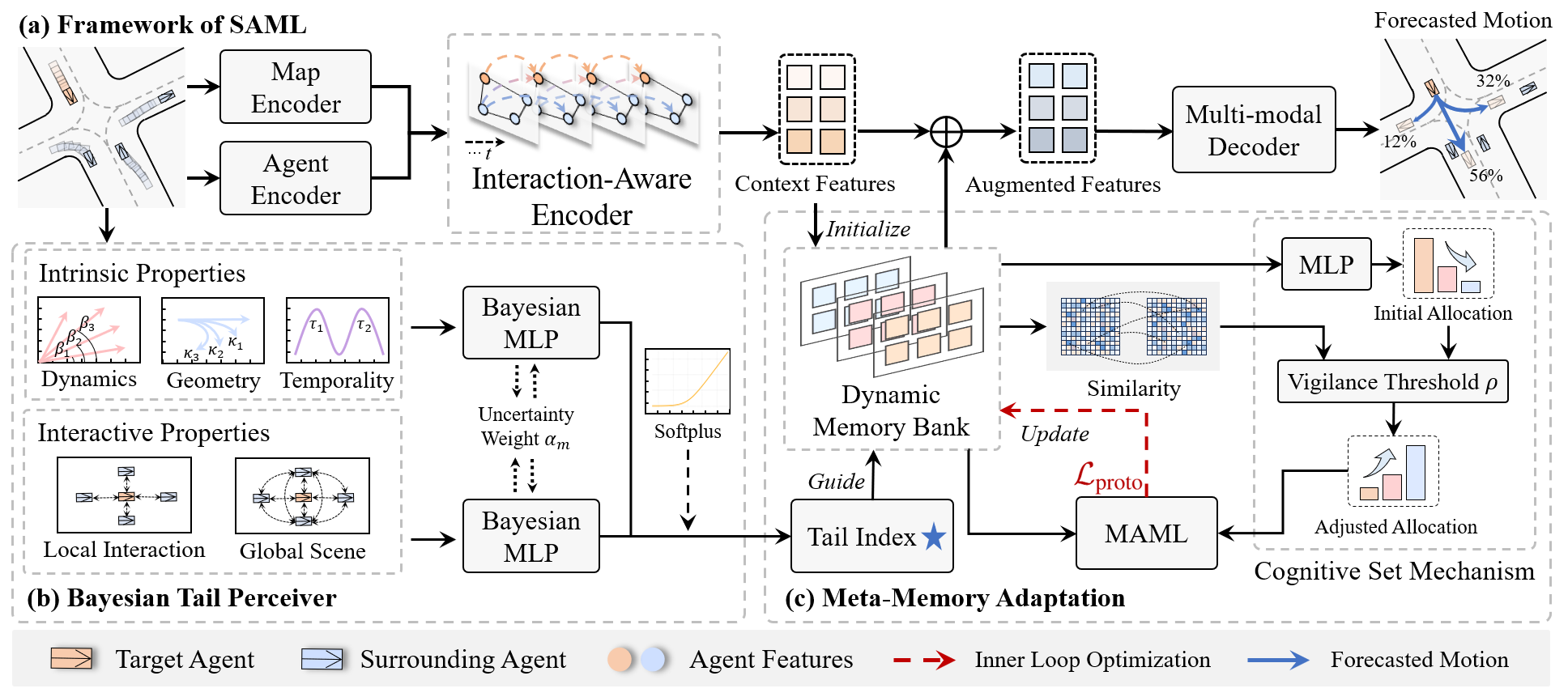}
   \vspace{-18pt}
\caption{Overview of the proposed SAML framework. 
(a) The overall model architecture. The model processes motion histories of a target agent and surrounding agents, along with HD map data, using four key modules: the Interaction-Aware Encoder, Bayesian Tail Perceiver, Meta-Memory Adaptation, and Multi-modal Decoder to generate a multimodal motion forecast.
(b) and (c) Detailed illustration of the Bayesian Tail Perceiver and the Meta-Memory Adaptation modules, respectively.}
\vspace{-12pt}
   \label{fig2}
\end{figure*}

\section{Related Work}
Motion forecasting, a core component of autonomous driving, is formulated as a time-series prediction problem. Early methods used Recurrent Neural Networks (RNNs) and variants like Gated Recurrent Units (GRUs) for temporal dependencies \cite{huang2021bayonet-GRU, liao2024bat-trans}. Social LSTM introduced “social forces” to model agent interactions \cite{alahi2016social, liao2024cognitive}, inspiring graph-based approaches where agents are nodes and interactions are edges \cite{mohamed2020social, xu2022adaptive-GNN, liao2024mftraj, wang2025nest}. Attention-based models later dominated, capturing non-local dependencies across agents and maps \cite{salzmann2020trajectron++, liao2025minds, liao2024cdstraj}. However, benchmark gains have plateaued, as routine success fails to ensure robustness in rare, safety-critical events, motivating long-tail research.

The long-tail phenomenon, studied in machine learning \cite{longtail-1}, is acute in autonomous driving due to imbalanced naturalistic data \cite{zhou2022long, zhang2024tract}. Routine behaviors dominate, while rare events like evasive maneuvers or intersections conflicts are underrepresented, biasing models toward common patterns and risking safety in long-tail cases \cite{li2023graph-risk}. Tackling this is essential for deployment.

Existing methods for long-tail forecasting can be grouped into data-centric and model-centric strategies:

1) Data-centric strategies rebalance datasets via oversampling rare events or undersampling frequent ones, risking overfitting or loss \cite{han2005borderline-resample, wang2025wake}. Advanced techniques generate synthetic samples with VAEs \cite{makansi2021on-exposing, wang2023fend}, GANs \cite{yang2024dynamic, liao2025cot}, or diffusion models \cite{bae2024singulartrajectory}, but they may add artifacts, require validation, and lack real-world guarantees.

2) Model-centric approaches enhance recognition without data changes, using loss re-weighting for rare samples \cite{ross2017focal-loss}, contrastive learning to distinguish behaviors such as Hi-SCL \cite{lan2024hi-scl} and AMD \cite{rao2025amd}, or meta-learning for few-shot adaptation \cite{li2024metatra}, though limited in multi-agent long-tail contexts.

While both data-centric and model-centric approaches provide partial solutions, there is still no framework that defines and adapts to long-tail events in a differentiable, semantically meaningful way—the motivation behind our proposed SAML framework.

\section{Methodology}
\subsection{Problem Formulation}
Given the motion histories of all agents in a scene $X = \{p, v, h\}$ and the static map context $M$, our objective is to forecast the future motion of a target agent $Y = \{p\}$. The history $X$ comprises sequences of kinematic state vectors, each encoding an agent's position $p$, velocity $v$, and heading angle $h$. To capture the multi-modal nature of future behavior, we formulate the task as learning a model to estimate the conditional probability distribution $P(Y | X, M)$. In practice, this distribution is approximated by a set of $K$ motion modes with associated confidence scores.

\subsection{Overall Framework and Motivation} 
SAML, illustrated in Figure~\ref{fig2}, jointly identifies and adapts to long-tail motions. An Interaction-Aware Encoder extracts scene features from motion histories and map context, while a Bayesian Tail Perceiver fuses intrinsic and interactive metrics into a continuous, uncertainty-aware Tail Index. Guided by this Tail Index, a meta-memory module with dynamic prototypes and a cognitive set mechanism emphasizes rare but safety-critical motions and mitigates bias toward common patterns. The adapted features are then decoded into multimodal motions, enabling SAML to handle long-tail cases efficiently even with limited data.

\subsection{Differentiable Definition of the Long Tail}
Existing methods for long-tail motion identification heavily rely on clustering techniques or model-specific forecasting errors. These approaches make it difficult to explain why a motion is long-tail, and error-dependent methods hinder end-to-end differentiable optimization. To address these issues, we define a motion's "tailness" by deconstructing long-tail events into semantically meaningful, fully differentiable metrics based on intrinsic and interactive properties. Specifically, we define three intrinsic properties—Kinematic Dynamism, Geometric Complexity, and Temporal Irregularity—and two interactive properties—Local Interaction Risk and Global Scene Risk. The conceptual definitions are as follows, with more details in the \textbf{Appendix A}.

\subsubsection{Intrinsic Properties}
Intrinsic properties quantify the inherent complexity of an agent's individual motion, analyzed along kinematic, geometric, and temporal dimensions.

\begin{definition}[Kinematic Dynamism]
Kinematic dynamism assesses the magnitude of changes in motion states, reflecting external forces or abrupt transitions such as emergency braking or sharp turns. Key metrics include velocity volatility $C_v$ and rotational instability $C_\alpha$:
\begin{equation}
    C_v =  \sqrt{\mathbb{E}_t\left[\left\|\frac{{v}(t) - {v}(t-\Delta t)}{\Delta t}\right\|_2^2\right]}
\end{equation}
\begin{equation}
    C_{\alpha} = \sqrt{\mathbb{E}_t\left[\left(\frac{\omega(t) - \omega(t-\Delta t)}{\Delta t}\right)^2\right]}
\end{equation}
where ${v}(t)$ is velocity, $\omega(t)$ is angular velocity. 
\end{definition}
Other metrics include acceleration instability $C_j$, heading volatility $C_\omega$, and movement direction volatility $C_{{vd}}$, with more details in the \textbf{Appendix}.

\begin{definition}[Geometric Complexity]
Geometric complexity measures the spatial curvature of the motion path, indicating turns or evasive maneuvers. A primary metric is curvature intensity $C_{\kappa}$:
\begin{equation}
       \kappa(t) = \frac{|v_x(t)a_y(t) - v_y(t)a_x(t)|}{[v_x(t)^2 + v_y(t)^2]^{3/2}}, \quad C_{\kappa} = \mathbb{E}_t[|\kappa(t)|]
\end{equation}
where $\kappa(t)$ is instantaneous curvature and $v_x(t)$, $v_y(t)$, $a_x(t)$, $a_y(t)$ are velocity and acceleration components. 
\end{definition}
Another metric is curvature volatility $C_{\Delta\kappa}$, with more details in the \textbf{Appendix A}.

\begin{definition}[Temporal Irregularity]
Temporal irregularity evaluates the predictability of velocity patterns through time-series analysis. It captures aperiodic fluctuations such as stop-and-go traffic by measuring changes in the velocity autocovariance function. The autocovariance fluctuation $C_{\Delta\gamma}$ is defined as:
\begin{equation} \label{eq:gamma_def}
    \gamma(\tau) = \mathbb{E}_t\left[ ({v}(t) - \bar{{v}}) \cdot ({v}(t+\tau) - \bar{{v}}) \right]
\end{equation}
\begin{equation} \label{eq:c_delta_gamma}
    C_{\Delta\gamma} = \frac{1}{T_h-1} \sum_{\tau=1}^{T_h-1} |\gamma(\tau) - \gamma(\tau-1)|
\end{equation}
where $\gamma(\tau)$ is the autocovariance function at time lag $\tau$, and $\bar{{v}}$ is the mean velocity over the observation window.
\end{definition}

\subsubsection{Interactive Properties}
Interactive properties quantify the potential risk and abnormality of a motion within a multi-agent context, evaluated at local and global levels.

\begin{definition}[Local Interaction Risk]
Local interaction risk assesses immediate threats from neighboring agents, indicating potential collisions in close proximity such as near-miss conflicts or aggressive tailgating. A Key metric is inverse time-to-collision (TTC) $R_{\text{ittc}}$, defined as:
\begin{equation}
    R_{\text{ittc}} = \mathbb{E}_t\left[ \max_{j \in \mathcal{A}_i} \left( \frac{\left[-({v}_j(t) - {v}_i(t)) \cdot ({p}_j(t) - {p}_i(t))\right]_+}{\| {p}_j(t) - {p}_i(t) \|_2^2 } \right) \right]
\end{equation}
where $[x]_+ = \max(0, x)$, ${v}_i(t), {p}_i(t)$ are velocity and position of agent $i$, and $\mathcal{A}_i$ is the set of neighboring agents to the target agent $i$.
\end{definition}
Details of this definition, including longitudinal and lateral risks extending the Responsibility-Sensitive Safety (RSS) framework, are provided in the \textbf{Appendix A}.

\begin{definition}[Global Scene Risk]
Global scene risk evaluates overall threat from collective agent dynamics, capturing environmental complexity and density such as in dense traffic or chaotic intersections. A key metric is multi-agent conflict $R_{\text{mac}}$:
\begin{equation}
    R_{\text{mac}} = \mathbb{E}_t \left[ \frac{2}{N(N-1)} \sum_{1 \le i < j \le N} \text{ITTC}_{ij}(t) \right]
\end{equation}
where $N$ is the number of agents, $\text{ITTC}_{ij}(t)$ is inverse TTC between agents $i$ and $j$. 
\end{definition}
Other metrics include agent density $R{\text{ad}}$ and neighborhood instability $R_{\text{ni}}$, with more details in the \textbf{Appendix A}.

\subsection{Bayesian Tail Perceiver}
The Bayesian Tail Perceiver aggregates semantic features into a continuous, differentiable Tail Index while modeling uncertainty to tackle sparsity and variability in long-tail scenarios. It employs a dual-path design to separately encode intrinsic and interactive properties into \(f_i\) and \(f_r\), minimizing feature interference and capturing rare kinematic patterns alongside multi-agent risks. The Bayesian formulation yields a smoothed Tail Index, emphasizes rare cases via elevated uncertainty, and ensures reliable gradients for end-to-end training with limited data. Each path is handled by a dedicated Bayesian MLP to produce latent representations, which are fused to compute the Tail Index.
\begin{equation}
{z}_{i} = {W}_{i}^{(2)} \sigma({W}_{i}^{(1)} {F}_i + {b}_{i}^{(1)}) + {b}_{i}^{(2)}
\end{equation}
\begin{equation}
{z}_{r} = {W}_{r}^{(2)} \sigma({W}_{r}^{(1)} {F}_r + {b}_{r}^{(1)}) + {b}_{r}^{(2)}
\end{equation}
where $\sigma$ is the ReLU activation function. The parameters for each path and layer, $\theta_m^{(l)} = \{W_m^{(l)}, b_m^{(l)}\}$ for $m \in \{i, r\}$ and $l \in \{1, 2\}$, are sampled from approximate posterior distributions $q(\theta_m^{(l)})$, parameterized as diagonal Gaussians.

To fuse the paths, we introduce an uncertainty-guided weighting based on the KL-divergence between posterior $q(\theta_m)$ and prior $p(\theta_m)$ distributions. The fusion weights $\alpha_m$ are computed as:
\begin{equation}
\alpha_m = \frac{\exp(\lambda \cdot \text{KL}(q(\theta_m)||p(\theta_m)))}{\sum_{n \in \{i, r\}} \exp(\lambda \cdot \text{KL}(q(\theta_n)||p(\theta_n)))}
\end{equation}
where $\lambda$ is a temperature parameter. The final Tail Index $TI$ is obtained by linearly combining the latent representations and applying a Softplus activation to ensure non-negativity:
\begin{equation}
TI = \sigma_{\text{sp}}\left( {w}_{o}^\top (\alpha_{i}{z}_{i} + \alpha_{r}{z}_{r}) + b_{o} \right)
\end{equation}
where $\sigma_{\text{sp}}(x) = \log(1+e^x)$ is the Softplus activation function, and ${w}_{o}$, $b_{o}$ are the learnable weight and bias of the output layer, respectively. 

\subsection{Interaction-Aware Encoder}

To generate rich, context-aware representations for motion histories, we design the Interaction-Aware Encoder. The process begins with independent encoding of scene elements: the target agent $f_t$, surrounding agents $f_n$, and map lanes $f_l$ are processed using separate GRUs, supplemented by a temporal Transformer for the target to capture long-range dependencies, yielding initial features $F_t$, $F_n$, and $F_l$.

Subsequently, agent interactions are modeled by constructing a graph $G_a$ from $\{F_t, F_n\}$ and applying self-attention $\mathcal{A}_{\text{self}}$ to produce intermediate interaction-aware features $G'_a$, which are aggregated into $F'_t$:
\begin{equation}
G'_a = \mathcal{A}_{\text{self}}(G_a + P_a), \quad F'_t = \mathbb{E}[G'_a]
\end{equation}
where $P_a$ is learnable positional encoding, and $\mathbb{E}$ denotes average pooling over $G'_a$. Multi-modal features are then decoded using learnable queries $Q$ through chained cross-attention $\mathcal{A}_{\text{cross}}$, first attending to the agent context $G'_a$ and then to the map context $F_l$:
\begin{equation}
F_m = \mathcal{A}_{\text{cross}}(\mathcal{A}_{\text{cross}}(Q + F'_t, G'_a + P'_a, G'_a), F_l + P_l, F_l)
\end{equation}
where $P'_a$ and $P_l$ are learnable positional encodings. The resulting $F_m$ provides context-aware, multi-modal features for subsequent meta-learning. Detailed steps and derivations are provided in \textbf{Appendix B}.

\begin{table*}[htbp]
  \centering
    \begin{tabular}{ c|cccccc } 
    \toprule
    Model & Venue & $\text{minADE}_{10}$ & $\text{minADE}_{5}$ & $\text{minFDE}_{5}$ & $\text{minFDE}_{1}$ & MR$_5$ \\
    \midrule
    Trajectron++ \cite{salzmann2020trajectron++} & ECCV  & 1.51  & 1.88 &  5.63 & 9.52  & 0.70 \\
    MultiPath \cite{chai2020multipath} & CoRL &  -     & 1.44 & 4.83  & 7.69  & 0.76 \\
    AgentFormer \cite{yuan2021agentformer} & ICCV &  1.31     & 1.59 & 3.14  & \underline{6.45}  & - \\
    Autobot \cite{girgislatent} & ICLR & \underline{1.03}  & 1.37 & 3.40  & 8.19     & 0.62 \\
    THOMAS \cite{gilles2022thomas} & ICLR & 1.04  & 1.33 & -  & {6.71}     & 0.55 \\
    PGP   \cite{deo2022multimodal} & CoRL & 1.03  & 1.30 & 2.52  & 7.17     & 0.61 \\
    GoHome \cite{gilles2022gohome}  & ICRA  & 1.15  & 1.42 & -  & 6.99  & 0.57 \\
    ContextVAE \cite{xu2023context}  & RAL  & -     & 1.59 & 3.28  & 8.24  & - \\
    EMSIN \cite{ren2024emsin} & TFS & 1.36  & 1.77 & 3.56  & 9.06  & 0.54 \\
    SeFlow \cite{zhang2024seflow}  & ECCV  & 1.04  & 1.38 & -  & 7.89  & 0.60 \\
    AMD \cite{rao2025amd} & ICCV & 1.06  & 1.23 & 2.43  & 6.99    & \underline{0.50} \\
    NEST  \cite{wang2025nest}  & AAAI  & -  & \underline{1.18} &  \underline{2.39} & 6.87  & 0.50 \\
    \midrule
    \rowcolor{green!10} \textbf{SAML (Ours)}  & - & \textbf{1.01}  & \textbf{1.18} & \textbf{2.34}  & \textbf{6.33} & \textbf{0.48} \\
    \bottomrule
    \end{tabular}
      \caption{Comparison of the performance of various models across all samples on nuScenes dataset. \textbf{Bold} and \underline{underlined} text represent the best and second-best results, respectively. Cases marked with ('-') indicate missing values.}
  \label{tab1}
\end{table*}

\begin{table}[h]
  \centering
  \footnotesize    
  \setlength{\tabcolsep}{0.75mm}
    \begin{tabular}{@{} c|c|cccccc @{}}
    \toprule
    \multicolumn{1}{c}{\multirow{2}[2]{*}{}}  & \multicolumn{1}{c|}{\multirow{2}[2]{*}{Model}} & \multicolumn{6}{c}{Forecasting Horizon (s)} \\
\cmidrule{3-8}       \multicolumn{1}{c}{}        &       & 1     & 2     & 3     & 4     & 5     & \multicolumn{1}{c}{AVG} \\
    \midrule
    \multirow{8}[2]{*}{\rotatebox{90}{NGSIM}} 
          & CF-LSTM \cite{xie2021congestion} & 0.55  & 1.10  & 1.78  & 2.73  & 3.82  & 2.00 \\
          & iNATran \cite{chen2022vehicle} & 0.39  & 0.96  & 1.61  & 2.42  & 3.43  & 1.76 \\
          & STDAN \cite{chen2022intention} & 0.39  & 0.96  & 1.62  & 2.51  & 3.65  & 1.83 \\
          & WSiP \cite{wang2023wsip}  & 0.56  & 1.23  & 2.05  & 3.08  & 4.34  & 2.25 \\
          & HiT  \cite{liao2025toward}  & 0.38  & 0.90  & 1.42  & 2.08  & 2.87  & 1.53 \\
          & DEMO \cite{wang2025dynamics}  & \underline{0.36}  & \underline{0.86}  & 1.48  & 2.10  & 2.88  & 1.54 \\
          & CITF \cite{liao2025minds}  & \textbf{0.30}  & \textbf{0.81}  & \underline{1.42}  & \underline{2.04}  & \underline{2.82}  & \underline{1.48} \\
          &  \cellcolor{green!10} \textbf{SAML (Ours)}   & \cellcolor{green!10} 0.39     & \cellcolor{green!10} 0.90     &  \cellcolor{green!10}  \textbf{1.36}   & \cellcolor{green!10} \textbf{1.81}     & \cellcolor{green!10}  \textbf{2.41}    & \cellcolor{green!10} \textbf{1.37}  \\
    \hline
    \multirow{8}[2]{*}{\rotatebox{90}{HighD}} 
          & CF-LSTM \cite{xie2021congestion} & 0.18  & 0.42  & 1.07  & 1.72  & 2.44  & 1.17 \\
          & STDAN \cite{chen2022intention} & 0.19  & 0.27  & 0.48  & 0.91  & 1.66  & 0.70 \\
          & DRBP \cite{gao2023dual} & 0.41  & 0.79  & 1.11  & 1.40  & 2.58  & 1.26 \\
          & WSiP \cite{wang2023wsip}  & 0.20  & 0.60  & 1.21  & 2.07  & 3.14  & 1.44 \\
          & HiT \cite{liao2025toward}   & 0.08  & 0.13  & 0.22  & 0.39  & 0.61  & 0.29 \\
          & DEMO \cite{wang2025dynamics}  & \underline{0.06}  & 0.14  & 0.25  & 0.44  & 0.70  & 0.32 \\
          & CITF \cite{liao2025minds}  & \textbf{0.04}  & \textbf{0.09}  & \underline{0.18}  & \underline{0.30}  & \underline{0.43}  & \underline{0.21} \\
          & \textbf{SAML (Ours)} \cellcolor{green!10} & \cellcolor{green!10} 0.08  &\cellcolor{green!10} \underline{0.12}  &\cellcolor{green!10} \textbf{0.15}  &\cellcolor{green!10} \textbf{0.21}  &\cellcolor{green!10} \textbf{0.33}  &\cellcolor{green!10} \textbf{0.18} \\
    \toprule
    \end{tabular}
  \caption{Comparison of model performance across all samples on NGSIM and HighD datasets. Metric: RMSE.}
  \label{tab2}%
\end{table}%

\subsection{Meta-Memory Adaptation}
The Meta-Memory Adaptation module enables SAML to handle sparse and evolving long-tail motion patterns. Guided by the Tail Index, it combines a dynamic prototype memory with an adaptive cognitive set mechanism to reduce bias toward frequent behaviors, and leverages a MAML framework to achieve rapid few-shot adaptation to rare or emerging motions.

\subsubsection{Cognitive Set Mechanism}
To leverage the dynamic memory bank $M$, which stores $C$ class-prototypes representing distinct motion categories as detailed in \textbf{Appendix C}, we design a mechanism to assess the relevance of each prototype to the current motion history. Relying solely on a data-driven gating network may induce a cognitive set, favoring frequent patterns and overlooking novel or long-tail events. To mitigate this bias, we propose a cognitive set mechanism that acts as a vigilance system. It computes a base category allocation distribution $g$ using a MLP on a concatenated input $h = [F_m, F_i, F_r, TI]$, and calculates the normalized similarity $s$ between feature $F_m$ and prototype bank $M$:
\begin{equation}
g = \phi_S(\phi_M(h))
\end{equation}
\begin{equation}
s = \frac{F_m \cdot M^\top}{\|F_m\|_2 \|M\|_2} \cdot \tau, \quad \tau > 0
\end{equation}
where $\phi_M$ denotes the MLP, $\phi_S$ denotes the softmax function, and $\tau$ is a temperature parameter.

We then introduces a learnable vigilance threshold $\rho \in \mathbb{R}$ which dynamically modulates the initial allocation based on the maximum similarity score. This produces an adjusted allocation $g'$ enabling the model to override its initial judgment when a strong match with a rare prototype is detected:
\begin{equation}
\lambda = \sigma(\gamma (\max\{s\} - \rho))
\end{equation}
\begin{equation}
g' = \lambda \cdot g + (1 - \lambda) \cdot b_{\text{tail}}
\end{equation}
where $\sigma$ denotes the sigmoid function, $\gamma$ controls the steepness of the transition, and $b_{\text{tail}}$ is a bias vector that amplifies long-tail categories. This process equips the model with adaptive vigilance, mitigating cognitive fixation and ensuring that novel patterns receive appropriate attention.

\begin{table*}[htbp]
  \centering
    \begin{tabular}{c|cccccc}
    \toprule
    Model & Top 1\% & Top 2\% & Top 3\% & Top 4\% & Top 5\% & All \\
    \midrule
    PGP \cite{deo2022multimodal}  & 8.86/21.92 & 7.21/17.90 & 6.24/15.68 & 5.52/13.77 & 5.02/12.44 & 1.28/2.52 \\
    Q-EANet \cite{chen2024q} & {7.55}/{18.78} & \underline{6.15}/\underline{15.58} & \underline{5.44}/\underline{13.76} & \underline{4.94}/{12.49} & \underline{4.55}/{11.49} & 1.20/{2.45} \\
    LAformer \cite{liu2024laformer} & 8.19/19.03 & 6.73/15.81 & 5.89/13.90 & 5.33/12.60 & 4.90/11.61 & 1.19/{2.42} \\
    UniTraj (MTR) \cite{feng2024unitraj} & 7.84/21.69 & 6.44/18.06 & 5.69/15.95 & 5.18/14.49 & 4.78/13.37 & \textbf{1.15}/2.61 \\
    AMD \cite{rao2025amd} & \underline{7.50}/\underline{18.47} & {6.37}/{15.71} & {5.65}/13.99 & 5.08/\underline{12.45} & 4.62/\underline{11.36} & {1.23}/\underline{2.39} \\
    \hline
   \rowcolor{green!10} \textbf{SAML (Ours)}  & \textbf{6.21}/\textbf{14.72} & \textbf{5.36}/\textbf{12.36} & \textbf{5.09}/\textbf{11.50} & \textbf{4.48}/\textbf{10.07} & \textbf{4.21}/\textbf{9.41} & \underline{1.18}/\textbf{2.34} \\
   \rowcolor{green!10} \textbf{SAML (Ours) (50\%)} & 7.81/19.33 & 6.48/16.02 & 5.75/14.07 & 5.13/12.57 & 4.73/11.52 & 1.30/2.56 \\
    \toprule
    \end{tabular}%
    \caption{Worst-case performance comparison on the nuScenes dataset, reported in minADE$_5$ / minFDE$_5$. The Top 1-5\% subsets are defined for each model individually based on its own worst-performing samples, ranked by minFDE$_5$.}
    \vspace{-9pt}
  \label{tab3}%
\end{table*}%

\subsubsection{MAML-driven Memory Adaptation}
To enable rapid adaptation to emerging long-tail distributions, we employ a MAML framework, optimizing memory prototypes for few-shot generalization in data-sparse scenarios. The prototype memory $M$ is refined using a contrastive loss to align features with class prototypes:
\begin{equation}
\mathcal{L}_{\text{proto}} = -\frac{1}{B} \sum_{i=1}^B \log \sigma(\sum_{k=1}^C g'_{i,k} s_{i,k} - \sum_{k=1}^C (1 - g'_{i,k}) s_{i,k})
\end{equation}
where $g'_{i,k}$ denotes the element for category $k$ in the adjusted category allocation $g'_i$ of sample $i$, and $s_{i,k}$ is the element for category $k$ in the similarity vector $s_i$ of sample $i$.

In the inner loop, $M$ is updated via gradient descent:
\begin{equation}
M' = M - \alpha \nabla_{M} \mathcal{L}_{\text{proto}}
\end{equation}
where $\alpha$ is the inner-loop learning rate. The outer loop optimizes the model parameters for cross-task generalization. This meta-learning approach ensures rapid alignment with long-tail patterns, such as sudden evasive actions, enhancing forecasting robustness. The refined memory $M'$ is then used to generate augmented features:
\begin{equation}
F_v = F_m + \sigma(\phi_M(h)) \cdot (g' \cdot M')
\end{equation}
where $F_v$ is the augmented feature, $\sigma$ denotes the sigmoid function, $\phi_M$ is a MLP, and $g'$ is the adjusted category allocation from the cognitive set mechanism.

\subsection{Multi-modal Decoder}

The Multi-modal decoder transforms the augmented features $F_v$ into future motion forecasts. We employ a GRU and MLP to generate multi-modal future motion representations, mapped to a Laplace distribution to capture uncertainty in long-tail scenarios. The Laplace distribution, with its peaked and heavy-tailed nature, is well-suited for modeling central tendencies and extreme deviations in long-tail motions. This approach produces diverse and robust motion forecasts, enhancing sensitivity to rare driving events. The model is trained end-to-end using a composite loss function, detailed in the \textbf{Appendix D}.

\section{Experiments}
\subsection{Experimental Setup}
We conduct a comprehensive evaluation of our framework on three diverse, large-scale datasets: the urban-centric nuScenes~\cite{caesar2020nuscenes}, and the highway-focused NGSIM~\cite{deo2018convolutional} and HighD~\cite{krajewski2018highd}. To ensure fair comparison, we adhere to the established evaluation protocol for each dataset. For the multi-modal nuScenes benchmark, we report the standard metrics of minADE$_K$, minFDE$_K$, and MR$_K$ over $K$ modes. For the NGSIM and HighD datasets, we report the RMSE. Further details on datasets, metric definitions, and implementation are provided in the \textbf{Appendix E}.

\subsection{Quantitative Results}
\subsubsection{Overall Performance Comparison} 
We evaluate the overall performance of our model against a range of state-of-the-art methods on three diverse datasets. The results on the urban-centric nuScenes benchmark, presented in Table~\ref{tab1}, demonstrate that our model achieves top performance across all metrics. Notably, our model achieves a 5.7\% error reduction in the minFDE$_1$ metric compared to the second-best method. Table~\ref{tab2} details the performance on the highway datasets NGSIM and HighD, where our model exhibits exceptional capabilities. It consistently outperforms all other models in long-term forecasting (3s-5s horizons) on both datasets, with a 14.5\% improvement over the second-best in 5s RMSE on NGSIM and a 23.3\% improvement on HighD. These results affirm the effectiveness and robustness of our proposed model in both urban and highway environments.

\subsubsection{Worst-Case Performance Comparison} 

Traditional long-tail evaluations often define hard cases based on the errors produced by a single, fixed baseline model, which can introduce inherent biases. To enable a more equitable comparison, we assess each model's performance in its respective worst-case scenarios. Specifically, for every model, we rank all test samples according to the minADE$_5$ errors it generates and select the top 1\% to 5\% of samples where that model exhibits the poorest performance. This protocol evaluates the upper bound of Forecasting error for each method. As illustrated in Table~\ref{tab3}, our proposed model exhibits superior performance. On the top 1\% of the most challenging samples, SAML achieves a minADE$_5$ of 6.21 m, representing a 17.2\% reduction relative to the second-best baseline. The advantage in minFDE$_5$ is even more substantial, with a 20.3\% reduction in error. These results indicate that SAML's worst-case performance is markedly superior to that of other state-of-the-art methods. Notably, our model variant trained on only 50\% of the data, SAML (50\%), still surpasses several fully trained baselines, such as LAformer and UniTraj (MTR), thereby demonstrating competitive efficacy even with reduced data volumes. This underscores the targeted and efficient enhancements to long-tail samples afforded by our semantic meta-learning framework.

\subsection{Efficiency Analysis}
To evaluate the efficiency of our proposed model, we conduct a comparative analysis of inference time and accuracy using the nuScenes dataset. All models are tested on a single NVIDIA RTX 3090 GPU. As presented in Table~\ref{tab4}, SAML demonstrates a distinct advantage in computational efficiency. The results indicate that our model achieves state-of-the-art accuracy while operating at a significantly faster inference speed of 21 ms compared to other high-performing methods, such as LAformer \cite{liu2024laformer}, making it a practical and effective solution for real-world deployment.

\begin{table}[h]
  \centering
  \small
  \setlength{\tabcolsep}{0.65mm}
    \begin{tabular}{c|ccc}
    \toprule
    Model & Time & $\text{minADE}_{5}$ & $\text{minFDE}_{1}$ \\
    \midrule
    Trajectron++ \cite{salzmann2020trajectron++} & \underline{38}    & 1.88  & 9.52 \\
    AgentFormer \cite{yuan2021agentformer} & 107   & 1.59  & \underline{6.45} \\
    PGP \cite{deo2022multimodal}  & 215   & 1.30   & 7.17 \\
    LAformer \cite{liu2024laformer} & 115   & \underline{1.19}  & 6.95 \\
    VisionTrap \cite{moon2024visiontrap} & 53    &   1.35    & 8.72 \\
    \hline
    \rowcolor{green!10} \textbf{SAML (Ours)} & \textbf{21} & \textbf{1.18} & \textbf{6.33} \\
    \toprule
    \end{tabular}%
    \caption{Efficiency comparison on the nuScenes dataset.}
    \vspace{-9pt}
  \label{tab4}%
\end{table}%

\begin{table}[htbp]
  \centering
  \footnotesize
  \setlength{\tabcolsep}{0.6mm}
    \begin{tabular}{c|ccccc}
    \toprule
    \multirow{2}[2]{*}{Comp.} & \multicolumn{5}{c}{Ablation Model} \\
\cmidrule{2-6}          & A     & B     & C     & D     & E \\
    \midrule
    BTP   & $\times$     & $\checkmark$     & $\checkmark$     & $\checkmark$     & $\checkmark$ \\
    IAE   & $\checkmark$     & $\times$     & $\checkmark$     & $\checkmark$     & $\checkmark$ \\
    CSM   & $\checkmark$     & $\checkmark$     & $\times$     & $\checkmark$     & $\checkmark$ \\
    MAML  & $\checkmark$     & $\checkmark$     & $\checkmark$     & $\times$     & $\checkmark$ \\
    \midrule
    Top 1\% & 7.87/19.17 & 7.86/19.18 & 7.89/19.21 & 7.39/18.15 & \textbf{6.21}/\textbf{14.72} \\
    Top 2\% & 6.29/15.51 & 6.45/15.72 & 6.47/15.78 & 6.22/15.30 & \textbf{5.36}/\textbf{12.36} \\
    Top 3\% & 5.34/13.43 & 5.62/13.73 & 5.84/14.16 & 5.58/13.75 & \textbf{5.09}/\textbf{11.50} \\
    All   & 1.26/2.46 & 1.36/2.64 & 1.33/2.56 & 1.30/2.52 & \textbf{1.18}/\textbf{2.34} \\
    \bottomrule
    \end{tabular}%
    \caption{Ablation results (minADE$_5$/minFDE$_5$) on nuScenes dataset. BTP: Bayesian Tail Perceiver; IAE: Interaction-Aware Encoder; CSM: Cognitive Set Mechanism; MAML: MAML-driven Memory Adaptation.}
    \vspace{-9pt}
  \label{tab5}%
\end{table}%

\subsection{Ablation Studies}

To validate the effectiveness of each key component in our framework, we conduct a series of ablation studies on the nuScenes dataset, with results presented in Table~\ref{tab5}. While the full model (Model E) establishes the best performance, the degradation observed in the ablated variants highlights the contribution of each module. Most notably, the removal of the Cognitive Set Mechanism (Model C) results in the most significant performance drop. Compared to our full model, this variant sees its minADE$_5$ and minFDE$_5$ on the Top 1\% set increase by 27.1\% and 30.5\%, respectively. This severe degradation underscores the critical role of the CSM in mitigating the model's inherent bias towards common patterns and enhancing its sensitivity to rare, long-tail events. Additionally, removing the Bayesian Tail Perceiver (Model A), replacing the Interaction-Aware Encoder (Model B), or ablating the MAML-driven adaptation (Model D) all lead to a noticeable decline in performance, particularly on the most challenging long-tail subsets. This confirms the necessity of our core components and demonstrates that they work synergistically to achieve the final performance.

\begin{figure}[h]
  \centering
  \vspace{-4pt}
  \includegraphics[width=1\linewidth]{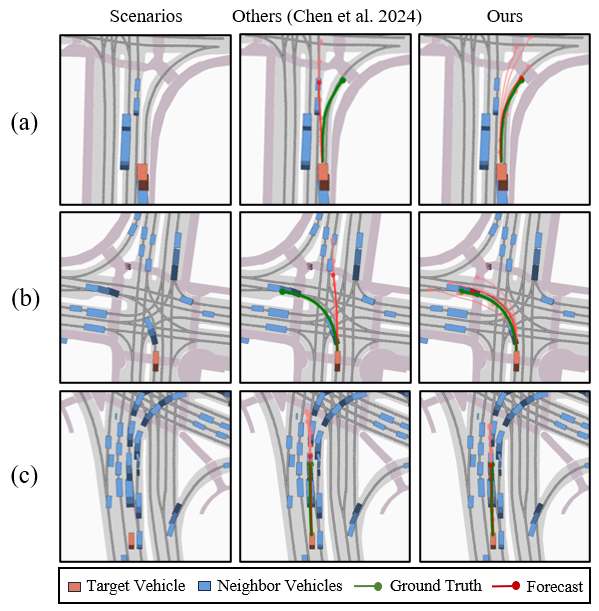}
   \caption{Visualization of long-tail performance on the nuScenes dataset. Red denotes the highest-probability forecast, and pink represents other multimodal options.}
   \vspace{-9pt}
   \label{fig3}
\end{figure}

\subsection{Qualitative Results}
\subsubsection{Long-tail Performance}

Figure~\ref{fig3} compares our SAML model with another model \cite{chen2024q} on multimodal motion forecasting in various long-tail scenarios from the nuScenes dataset. Panels (a), (b), and (c) illustrate right-turn, left-turn, and deceleration maneuvers in congested intersections, respectively. The other model, biased towards common straight-line patterns, struggles with abrupt changes, leading to deviations from ground truth and limited multimodal diversity. In contrast, SAML achieves excellent performance in these long-tail scenarios. In panel (a), the target motion exhibits high velocity volatility (\textbf{top 10\%} in kinematic dynamism), an intrinsic property reflecting abrupt deceleration, enabling SAML to forecast the right turn precisely by learning the deceleration feature preceding right turns. Panel (b) shows elevated lateral risk (\textbf{top 1\%} in local interactive risk), arising from numerous conflict points typical in left turns at intersections, while panel (c) features high scene density and longitudinal risk (\textbf{top 3\%} in local interactive risk), due to dense traffic impeding forward motion. SAML effectively learns these distinct long-tail features through its semantic-aware meta-learning framework, achieving precise forecasting. Additional visualizations of long-tail scenarios are provided in \textbf{Appendix F}.

\subsubsection{Failure Cases}
\begin{figure}[h]
  \centering
  \vspace{-4pt}
  \includegraphics[width=1\linewidth]{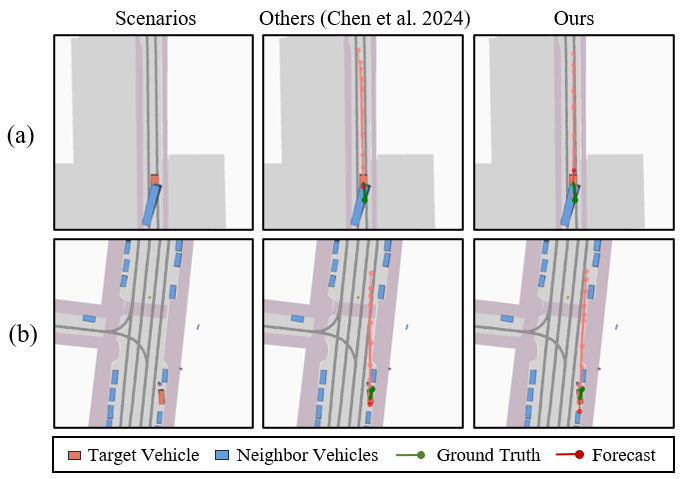}
   \caption{Visualization of failure cases on the nuScenes dataset. Red denotes the highest-probability forecast, and pink represents other multimodal options.}
   \vspace{-9pt}
   \label{fig4}
\end{figure}

To investigate the limitations of SAML, we analyze two contrasting failure cases shown in Figure~\ref{fig4}, which highlight a deeper challenge: resolving ambiguity within extreme long-tail scenarios. In the panel (a), a vehicle unexpectedly reverses on a two-way road—a highly rare maneuver—that SAML fails to anticipate, instead defaulting to a high-probability forward motion forecast. Conversely (b), another vehicle is angled in a manner suggestive of backing into a parking spot; SAML detects the anomalous orientation and forecasts a reversal, yet the vehicle moves forward slightly to adjust its position. These cases collectively illustrate that while SAML can identify a scenario's rarity (its "tailness"), it struggles to disambiguate the driver's intent when the cues from these rare events are contradictory. This limitation highlights the need for future research to focus on resolving the semantic ambiguity arising from conflicting cues, a defining challenge of extreme long-tail events.

\section{Conclusion}
In this paper, we introduce SAML, a novel framework that features a differentiable semantic meta-learning approach for long-tail motion forecasting in autonomous driving. Our method establishes the first principled way to quantify a motion's tailness, enabling interpretable, end-to-end optimization for rare events. The effectiveness of SAML is validated through extensive experiments, demonstrating superior performance in diverse long-tail scenarios. Furthermore, our analysis of failure cases reveals a deeper challenge beyond simple rarity detection. We find that while SAML excels at identifying a scenario's tailness, it struggles to resolve the ambiguity of contradictory cues within extreme events. This limitation highlights the need for future research to focus on resolving the semantic ambiguity arising from conflicting cues—a defining challenge of the field.

\section{Acknowledgments}
This work was supported by the Science and Technology Development Fund of Macau [0129/2022/A, 0122/2024/RIB2, 0215/2024/AGJ, 001/2024/SKL], the Research Services and Knowledge Transfer Office, University of Macau [SRG2023-00037-IOTSC, MYRG-GRG2024-00284-IOTSC], the Shenzhen-Hong Kong-Macau Science and Technology Program Category C [SGDX20230821095159012], the Science and Technology Planning Project of Guangdong [2025A0505010016], National Natural Science Foundation of China [52572354], the State Key Lab of Intelligent Transportation System [2024-B001], and the Jiangsu Provincial Science and Technology Program [BZ2024055].

\bibliography{aaai2026}

\clearpage
\appendix
\twocolumn[
  \begin{center}
    \Huge\bfseries \appendixpagename 
  \end{center}
  \vspace{1em}
]
\setcounter{equation}{0}
\setcounter{figure}{0}

\section{A. Differentiable Long-Tail Quantification}
To overcome the ambiguity of indirect, model-dependent definitions based on forecasting error, we introduce a principled, semantics-based, differentiable framework for quantifying long-tail motions. This framework aims to capture intrinsic complexity directly from the motion data, providing a model-agnostic and semantically meaningful metric. We posit that the long-tail nature of a motion arises from two orthogonal yet complementary aspects: its Intrinsic Properties, which describe the complexity of the agent's individual motion, and its Interactive Properties, which quantify the risk and abnormality of its behavior within a multi-agent context. By modeling both aspects, we can comprehensively quantify the degree to which a motion is long-tail.

\subsection{Intrinsic Properties} 
A motion, as a time series of an agent's state, exhibits intrinsic properties that reveal its behavioral complexity, which is crucial for identifying long-tail scenarios. To this end, we establish a principled system that deconstructs a motion's intrinsic properties along fundamental physical and mathematical dimensions: \textbf{Kinematic Dynamism}, which captures the severity of changes in motion states over time; \textbf{Geometric Complexity}, which quantifies the intricacy of the motion's spatial form; and \textbf{Temporal Irregularity}, which measures the temporal regularity and regularity of its motion patterns. This structured approach allows us to derive a set of specific, differentiable metrics to comprehensively assess the intrinsic complexity of a motion.

\setcounter{definition}{0}
\begin{definition}[Kinematic Dynamism]
\end{definition}

The kinematic dynamism of a motion quantifies the severity of changes in an agent's motion state over time. In classical mechanics, significant variations in higher-order temporal derivatives of motion often correspond to the agent being subjected to substantial external forces or undergoing abrupt internal state transitions. In motion forecasting, this directly maps to typical long-tail scenarios such as emergency braking, sharp turns, or evasive maneuvers. Given the historical motion of the target agent $\mathbf{X}^{obs}$, we define the following metrics to measure kinematic dynamism from both translational and rotational perspectives. In these definitions, $\mathbb{E}_t[\cdot]$ denotes the mean taken over all time steps within the observation window.

\begin{itemize}
    \item \textbf{Velocity Volatility ($C_v$)}: Quantifies the degree of change in translational velocity ${v}(t)$. It is calculated as the Root Mean Square (RMS) of the magnitude of the instantaneous acceleration ${a}(t)$. This metric is particularly sensitive to unstable motion states, which are a common cause of motion forecasting model failures.
    \begin{equation}
        C_v = \sqrt{\mathbb{E}_t[\|{a}(t)\|_2^2]} =  \sqrt{\mathbb{E}_t\left[\left\|\frac{{v}(t) - {v}(t-\Delta t)}{\Delta t}\right\|_2^2\right]}
    \end{equation}

    \item \textbf{Acceleration Instability ($C_{{j}}$)}: Measures the severity of changes in acceleration, a quantity known as jerk. Its magnitude serves as a hallmark of extreme driving behaviors, such as emergency evasions, and is critical for comfort and safety assessment.
    \begin{equation}
        C_{{j}} = \sqrt{\mathbb{E}_t\left[\left\|\frac{{a}(t) - {a}(t-\Delta t)}{\Delta t}\right\|_2^2\right]}
    \end{equation}

    \item \textbf{Heading Volatility ($C_{\omega}$)}: Quantifies the rate of change of the heading angle $\theta(t)$, which is the angular velocity $\omega(t)$. Elevated values of $C_{\omega}$ are directly associated with sharp turns or frequent lane changes, posing significant challenges for forecasting in complex areas like intersections.
    \begin{equation}
        C_{\omega} = \sqrt{\mathbb{E}_t[\omega(t)^2]} = \sqrt{\mathbb{E}_t\left[\left(\frac{\theta(t) - \theta(t-\Delta t)}{\Delta t}\right)^2\right]}
    \end{equation}

    \item \textbf{Rotational Instability ($C_{\alpha}$)}: Measures the rate of change of angular velocity $\omega(t)$, which corresponds to the angular acceleration. This metric captures abrupt rotational movements, making it a key indicator for high-risk scenarios that require rapid heading correction.
    \begin{equation}
        C_{\alpha} = \sqrt{\mathbb{E}_t\left[\left(\frac{\omega(t) - \omega(t-\Delta t)}{\Delta t}\right)^2\right]}
    \end{equation}

    \item \textbf{Movement Direction Volatility ($C_{{vd}}$)}: Quantifies the rate of change of the actual movement direction angle $\phi(t) = \operatorname{arctan2}(v_y(t), v_x(t))$. It is crucial for identifying behaviors where heading and movement direction diverge, for instance, in vehicle skids or pedestrian detours.
    \begin{equation}
        C_{{vd}} = \sqrt{\mathbb{E}_t\left[\left(\frac{\phi(t) - \phi(t-\Delta t)}{\Delta t}\right)^2\right]}
    \end{equation}
\end{itemize}

\begin{definition}[Geometric Complexity]
\end{definition}
As a sequence of points in space, a motion's geometric form is fundamental to understanding an agent's behavior and anticipating its future path. To quantify the spatial intricacy of a motion, we introduce the concept of geometric complexity. This is primarily achieved by analyzing the motion's local bending characteristics and their rate of change. We define the following metrics:

\begin{itemize}
    \item \textbf{Curvature Intensity ($C_{\kappa}$)}: Measures the average degree of bending along the motion. This is quantified by the expected magnitude of the instantaneous curvature $\kappa(t)$. The value of $C_{\kappa}$ directly reflects the presence of significant turns or evasive maneuvers within the path.
    \begin{equation}
       \kappa(t) = \frac{|v_x(t)a_y(t) - v_y(t)a_x(t)|}{[v_x(t)^2 + v_y(t)^2]^{3/2}}, \quad C_{\kappa} = \mathbb{E}_t[|\kappa(t)|]
    \end{equation}
    where $v_x(t), v_y(t)$ are the longitudinal and lateral velocity components, and $a_x(t), a_y(t)$ are the longitudinal and lateral acceleration components, respectively.
    
    \item \textbf{Curvature Volatility ($C_{\Delta\kappa}$)}: Captures the non-smoothness of the motion's path by measuring how abruptly its curvature changes. This is evaluated by the expected magnitude of the curvature's rate of change, $\Delta{\kappa}(t)$. This metric is particularly sensitive to complex maneuvers, such as a sudden transition from a straight line to a sharp curve.
    \begin{equation}
        C_{\Delta\kappa} = \mathbb{E}_t[|\Delta\kappa(t)|] = \mathbb{E}_t[|\frac{\kappa(t) - \kappa(t-\Delta t)}{\Delta t}|]
    \end{equation}
\end{itemize}

\begin{definition}[Temporal Irregularity]
\end{definition}

In addition to its kinematic and geometric characteristics, we analyze the motion from the \textbf{temporal domain}. This dimension offers critical insights into a motion's complexity by examining its structure as an entire time series, with a focus on properties such as periodicity and temporal regularity. Irregular temporal structures, for instance, the erratic stop-and-go patterns in dense traffic, are hallmarks of long-tail scenarios. Drawing from time-series analysis, we formalize this concept by measuring the fluctuation of the velocity's autocovariance function.

\begin{itemize}
    \item \textbf{Autocovariance Fluctuation ($C_{\Delta\gamma}$)}: Measures the average absolute change in the velocity autocovariance function across different time lags to capture temporal irregularity. A high $C_{\Delta\gamma}$ indicates aperiodic or unstable velocity patterns, for instance, sudden braking at an intersection, which are hallmarks of long-tail motions.

    \begin{equation} \label{eq:gamma_def}
        \gamma(\tau) = \mathbb{E}_t\left[ ({v}(t) - \bar{{v}}) \cdot ({v}(t+\tau) - \bar{{v}}) \right]
    \end{equation}
    \begin{equation} \label{eq:c_delta_gamma}
        C_{\Delta\gamma} = \frac{1}{T_h-1} \sum_{\tau=1}^{T_h-1} |\gamma(\tau) - \gamma(\tau-1)|
    \end{equation}
where $\gamma(\tau)$ is the autocovariance function at time lag $\tau$, and $\bar{{v}}$ is the mean velocity over the observation window.

\end{itemize}

\subsection{Interactive Properties}
Motions in traffic scenes are not isolated time-series data; the coexistence of multiple social agents fundamentally shapes them within a shared physical space. Thus, analyzing the intrinsic physical properties of a single motion is insufficient to capture its behavioral complexity, especially in long-tail scenarios. We argue that a motion's deeper semantics are rooted in its interactivity: how an agent perceives, anticipates, and influences the actions of other road users. Abnormalities in these interactions, such as close-proximity cut-ins, risky negotiations, or a failure to react to potential conflicts, are the core drivers of high-risk events and forecast failures. Consequently, we approach the characterization of these critical interactive dynamics from the perspective of risk quantification. To this end, we systematically evaluate these safety-critical behaviors at two hierarchical levels: \textbf{Local Interaction Risk} and \textbf{Global Scene Risk}.

\begin{definition}[Local Interaction Risk]
\end{definition}

Local interaction risk aims to quantify the direct threat of collision between the target agent and a single other agent in the scene. Such local interactions form the basic building blocks of many long-tail scenarios, for instance, near-miss conflicts at intersections or aggressive tailgating on highways. By evaluating these pairwise dynamics, we capture the immediate risks that can lead to safety-critical events and pose significant challenges to forecasting models. We propose the following metrics:

\begin{itemize}
    \item \textbf{Inverse Time-to-Collision (ITTC)}: This metric quantifies risk from a temporal dimension by measuring the reciprocal of the Time-to-Collision (TTC), directly reflecting the imminence and urgency of a potential collision. The $R_{\text{ITTC}}$ between the target agent $i$ and a neighbor $j$ is defined as:
    \begin{footnotesize}
    \begin{equation}
        R_{\text{ITTC}} = \mathbb{E}_t\left[ \max_{j \in \mathcal{A}_i} \left( \frac{\left[-({v}_j(t) - {v}_i(t)) \cdot ({p}_j(t) - {p}_i(t))\right]_+}{\| {p}_j(t) - {p}_i(t) \|_2^2 } \right) \right]
    \end{equation}
    \end{footnotesize}
where $[x]_+ = \max(0, x)$, $\mathcal{A}_i$ is the set of neighboring agents to the target agent $i$, ${p}_i(t)$ and ${p}_j(t)$ are the positions of the agents.

    \item \textbf{Longitudinal Risk ($R_{\text{lon}}$)}: We introduce the concept of longitudinal risk from the Responsibility-Sensitive Safety (RSS) model~\cite{rss_ref}. This metric quantifies risk from a spatial dimension by first calculating a minimum safe distance and then mapping the deficit between this safe distance and the actual distance to a continuous risk value. The minimum required longitudinal separation $d_{x}(t, j)$ is defined for each neighbor $j$ as:
    \begin{equation} \label{eq:d_min_lon}
    \scriptsize
        {d_{x}(t, j) = \bigg[ v_i(t)\rho + \frac{1}{2}a_{\text{max}}\rho^2} + \frac{(v_i(t) + \rho a_{\text{max}})^2}{2b_{\text{min}}} - \frac{(v_j^{\text{veh}}(t))^2}{2b_{\text{max}}} \bigg]_{+}
    \end{equation}
    where the term $v_j^{\text{veh}}(t)$ is the velocity of neighbor $j$ if it is a vehicle, and is treated as zero otherwise. The terms $v_i(t)$ are the longitudinal velocities; $\rho$ is the reaction time; $a_{\text{max}}$ is the maximum acceleration; and $b_{\text{max}}$ is the maximum braking deceleration. The overall longitudinal risk is then computed as:
    \begin{equation} \label{eq:lon_risk}
    \small
        {R_{\text{lon}} = \mathbb{E}_t\left[ \max_{j} \left( 1 - \left(1 + \frac{[d_{x}(t, j) - d_{\text{lon}}(t, j)]_+}{\beta d_{x}(t, j)}\right)^{-\alpha} \right) \right]}
    \end{equation}
    where $d_{\text{lon}}(t, j)$ is the actual longitudinal distance, $\alpha, \beta$ are parameters governing the risk function's sensitivity.

    \item \textbf{Lateral Risk ($R_{\text{lat}}$)}: We introduce the concept of lateral risk, also derived from the RSS model, to assess spatial risk during lateral maneuvers such as lane changes or merges. It quantifies the risk by comparing the actual lateral separation to a formally defined minimum safe lateral distance. The minimum required lateral separation $d_{y}(t, j)$ is defined as:
    \begin{equation} \label{eq:d_min_lat}
    \small
    \begin{split}
        d_{y}(t, j) = \mu' &+ \bigg[ \left( \frac{v'_{i}(t) + v'_{i,\rho}(t)}{2}\rho + \frac{(v'_{i,\rho}(t))^2}{2b'_{\text{min}}} \right) \\
        & - \left( \frac{v'_{j}(t) + v'_{j,\rho_{\text{eff}}}(t)}{2}\rho_{\text{eff}} - \frac{(v'^{\text{veh}}_{j,\rho}(t))^2}{2b'_{\text{min}}} \right) \bigg]_{+}
    \end{split}
    \end{equation}
    where the effective reaction time for the neighbor $\rho_{\text{eff}}$ is $\rho$ for vehicles and $\rho_{\text{ped}}$ for pedestrians. The velocities after reaction time are $v'_{i,\rho}(t) = v'_{i}(t) + a'_{\text{max}}\rho$ and $v'_{j,\rho_{\text{eff}}}(t) = v'_{j}(t) - a'_{\text{max}}\rho_{\text{eff}}$. The terms $v'_{i}(t), v'_{j}(t)$ are the lateral velocities. The term $v'^{\text{veh}}_{j, \rho}(t)$ is the velocity of neighbor $j$ if it is a vehicle, and is treated as zero otherwise. $a'_{\text{max}}$ is the maximum lateral acceleration, $b'_{\text{min}}$ is the minimum plausible lateral deceleration, and $\mu'$ is a fixed safety margin. The lateral risk is then computed as:
    \begin{equation} \label{eq:lat_risk}
    \small
        R_{\text{lat}} = \mathbb{E}_t\left[ \max_{j} \left( 1 - \left(1 + \frac{[d_{y}(t, j) - d_{\text{lat}}(t, j)]_+}{\beta' d_{y}(t, j)}\right)^{-\alpha'} \right) \right]
    \end{equation}

    where $d'_{\text{lat}}(t, j)$ is the actual lateral distance, and $\alpha', \beta'$ are parameters governing the risk function's sensitivity.

\end{itemize}

\begin{definition}[Global Scene Risk]
\end{definition}

While local interaction risk effectively captures threats between pairs of agents, the risk in many long-tail scenarios extends beyond pairwise interactions. In complex situations, such as dense traffic or chaotic intersections, the dominant risk stems from the collective effect of all agents and their interplay---a property we term global scene risk. This dimension evaluates the overall environmental instability and risk profile, which we quantify using the following metrics:

\begin{itemize}
    \item \textbf{Multi-Agent Conflict ($R_{\text{mac}}$)}: This metric quantifies the overall conflict level of the entire scene by aggregating the Inverse Time-to-Collision (ITTC) over all unique pairs of agents. It serves as a measure of the scene's total interaction urgency.
    \begin{equation}
        R_{\text{mac}} = \mathbb{E}_t \left[ \frac{2}{N(N-1)} \sum_{1 \le i < j \le N} \text{ITTC}_{ij}(t) \right]
    \end{equation}
    where $N$ is the total number of agents in the scene, and $\text{ITTC}_{ij}(t)$ is the Inverse Time-to-Collision between agent $i$ and agent $j$ at time $t$.

    \item \textbf{Agent Density ($R_{\text{ad}}$)}: This metric measures the agent density within a specific radius $R$ around the target agent, directly reflecting the scene's crowdedness. Such dense environments are characteristic of urban long-tail scenarios, often constraining an agent's maneuvering space and increasing interaction complexity.
    \begin{equation}
        R_\text{ad} = \mathbb{E}_t \left[ \frac{1}{\pi R^2} \sum_{j \in \mathcal{A}_i} \mathbb{I}(\| {p}_i(t) - {p}_j(t) \|_2 \le R) \right]
    \end{equation}
    where $\mathcal{A}_i$ is the set of neighboring agents to the target agent $i$, ${p}_i(t)$ and ${p}_j(t)$ are the positions of the agents, and $\mathbb{I}(\cdot)$ is the indicator function which equals 1 if the condition is true, and 0 otherwise.

    \item \textbf{Neighborhood Instability ($R_{\text{ni}}$)}: This metric captures the collective volatility of the surrounding agents' motion by averaging their individual Velocity Volatility ($C_v$). A high value for $R_{\text{ni}}$ signifies a collectively unstable environment, which directly elevates the difficulty of forecasting the target agent's behavior.
    \begin{equation}
        R_\text{ni} = \mathbb{E}_{j \in \mathcal{A}_i}[C_v(j)]
    \end{equation}
    where $\mathcal{A}_i$ is the set of neighboring agents to the target agent $i$, and $C_v(j)$ is the Velocity Volatility of neighbor $j$, as computed over the observation horizon according to Definition 1.
\end{itemize}

\section{B. Interaction-Aware Encoder}
To generate rich, context-aware representations for motion histories, we design an Interaction-Aware Encoder. Initially, scene elements are encoded independently. The target agent's motion history $f_t \in \mathbb{R}^{T \times D}$ is processed by a Gated Recurrent Unit (GRU) followed by a temporal Transformer layer to capture long-range dependencies:
\begin{equation}
F_t = \phi_{Tr}(\phi_{GRU}(f_t))
\end{equation}
where $\phi_{GRU}$ denotes the GRU encoder, and $\phi_{Tr}$ is the Transformer layer.

Surrounding agents $f_n \in \mathbb{R}^{N \times T \times D}$ and map lanes $f_l \in \mathbb{R}^{L \times D}$ are encoded using separate GRUs:
\begin{equation}
F_n = \phi_{GRU}(f_n), \quad F_l = \phi_{GRU}(f_l)
\end{equation}

To model mutual influences among agents, we construct an agent graph $G_a$ by concatenating agent feature maps $\{F_t, F_n\}$. A self-attention mechanism is applied to capture global agent-agent interactions, producing interaction-aware features $G'_a$:
\begin{align}
G'_a &= \mathcal{A}_{\text{self}}(G_a + P_a) \\
&= \text{softmax}\left(\frac{G_a (G_a + P_a)^T}{\sqrt{d_k}}\right) G_a
\end{align}
where $P_a$ is a learnable positional encoding, and $d_k$ is the key dimension. These are aggregated into a target-centric graph representation $F'_t$:
\begin{equation}
F'_t = \mathbb{E}_{v \in G'_a}[v]
\end{equation}
where $\mathbb{E}_{v \in G'_a}[v]$ is average pooling over node features in $G'_a$.

We then employ $K$ learnable mode queries $Q$ to decode these contextual features into distinct future motion possibilities. These queries are refined through a two-stage cross-attention mechanism $\mathcal{A}_{\text{cross}}$, using $F'_t$ to enhance context-awareness. The queries first attend to the agent-interaction context, yielding socially-aware features $F_a$, which then attend to the map context to align motion intents with road geometry, producing multi-modal output features $F_m$:
\begin{align}
    F_a &=  \mathcal{A}_{\text{cross}}(Q + F'_t, G'_a + P'_a, G'_a) \\
   & =  \text{softmax}\left(\frac{(Q + F'_t) (G'_a + P'_a)^T}{\sqrt{d_k}}\right) G'_a
\end{align}
\begin{align}
    F_m &= \mathcal{A}_{\text{cross}}(F_a, F_l + P_l, F_l) \\
    &= \text{softmax}\left(\frac{F_a (F_l + P_l)^T}{\sqrt{d_k}}\right) F_l
\end{align}
where $P'_a$ and $P_l$ are learnable positional encodings.
The output $F_m$ provides a set of context-aware, multi-modal features for the subsequent meta-learning.

\section{C. Meta-Memory Adaptation}

\subsubsection{Dynamic Memory Bank}
To provide a structured, evolving knowledge base of motion patterns, we introduce a dynamic memory bank $M \in \mathbb{R}^{C \times D}$, storing $C$ class-prototypes. Each prototype represents a distinct motion category, defined by partitioning motion histories based on percentiles of their Tail Index. The prototype for category $c$, $M_c$, is initialized as the mean feature of the class and refined during training via a $TI$-weighted momentum update:
\begin{equation}
    M_c \leftarrow \eta M_c + (1 - \eta) \cdot \frac{\sum_{f \in \mathcal{B}_c} \exp({TI}(f)) \cdot f}{\sum_{f \in \mathcal{B}_c} \exp({TI}(f))}
\end{equation}
where $f$ is the feature vector of a sample, $\eta$ is a momentum parameter, and $\mathcal{B}_c$ is a mini-batch of samples from category $c$. This evolving memory bank $M$ provides a rich set of prototypes for subsequent feature augmentation.

\section{D. Training Loss}
\label{app:loss_details}

Our model is trained end-to-end using a composite loss function $L_{\text{total}}$ designed to holistically address the challenges of long-tail motion forecasting. This objective intricately balances a primary multi-modal forecasting loss $L_{\text{task}}$ for forecasting accuracy, with two specialized components aimed at long-tail learning. The first is a novel rank-based supervision loss $L_{\text{ti}}$ which guides our differentiable Tail Index to correlate with actual forecasting difficulty. This is complemented by a meta-learning loss $L_{\text{meta}}$ which promotes feature separation and rapid adaptation for robust generalization. The total loss is a weighted combination of these objectives:
\begin{equation} \label{eq:total_loss}
    L_{\text{total}} = L_{\text{task}} + \lambda_{1} L_{\text{ti}} + \lambda_{2} L_{\text{meta}}
\end{equation}
where $\lambda_{1}$ and $\lambda_{2}$ are hyperparameters balancing the different learning signals. 

\subsubsection{Task Loss ($L_{\text{task}}$)}
Our multi-modal forecasting task is optimized using a loss that combines regression and classification objectives. For a set of $K$ forecasted motions $\{\hat{Y}_k\}_{k=1}^K$ and their corresponding probabilities $\{p_k\}_{k=1}^K$, we first identify the best-matching motion, $\hat{Y}_{k^*}$, by finding the mode with the minimum Average Displacement Error (ADE) to the ground truth $Y_{gt}$. The task loss is then the negative log-likelihood (NLL) of this optimal mode:
\begin{equation}
    L_{\text{task}} = \mathcal{L}_{\text{reg}}(\hat{Y}_{k^*}, Y_{gt}) + \lambda_{\text{cls}} \cdot (-\log(p_{k^*}))
\end{equation}
where $\mathcal{L}_{\text{reg}}$ is the Mean Squared Error (MSE) between the optimal forecast and the ground truth, and $\lambda_{\text{cls}}$ is a balancing hyperparameter.

\subsubsection{Tail Index Supervision Loss ($L_{\text{ti}}$)}
To ensure our differentiable tail index is a meaningful measure of a motion's "tailness," we supervise it to learn the rank correlation with the forecasting error. This is achieved through a self-supervised loss that is robust to the absolute scale of the error. Within each batch of size $B$, we sort both the Tail Index values and their corresponding final ADE values. The loss is then the 1-Wasserstein distance between these two sorted distributions:
\begin{equation}
    L_{\text{ti}} = \frac{1}{B} \sum_{i=1}^{B} |{TI}_{\text{sorted}}^{(i)} - \text{ADE}_{\text{sorted}}^{(i)}|
\end{equation}
This encourages a monotonic relationship between the tail index and the actual forecasting difficulty.

\subsubsection{Meta-Learning Loss ($L_{\text{meta}}$)}
The meta-learning loss, $L_{\text{meta}}$, enhances generalization to rare scenarios by structuring the feature embedding space. It consists of two main components:
\begin{itemize}
    \item \textbf{Feature Separation Loss ($L_{\text{sep}}$)}: This contrastive-style loss encourages features from different long-tail categories to be separable in the embedding space.
    \item \textbf{MAML-based Loss ($L_{\text{maml}}$)}: This component, inspired by Model-Agnostic Meta-Learning (MAML), optimizes the model's parameters for rapid adaptation to new data distributions with few examples.
\end{itemize}
The total meta-learning loss is a weighted sum of these two objectives:
\begin{equation}
    L_{\text{meta}} = w_{s} L_{\text{sep}} + w_{m} L_{\text{maml}}
\end{equation}
where $w_{s}$ and $w_{m}$ are balancing hyperparameters.

\section{E. Experiments Stetup}
\subsection{E.1 Datasets}
To thoroughly evaluate the performance of our proposed framework across diverse and challenging conditions, we conducted extensive experiments on three large-scale, publicly available datasets. Below is a detailed overview of each dataset and our data processing methodology.

\begin{itemize}
    \item \textbf{nuScenes}: The nuScenes dataset~\cite{caesar2020nuscenes} is a comprehensive, large-scale benchmark for urban autonomous driving, with data collected in Boston and Singapore. It consists of 1000 scenes, each 20 seconds in duration, annotated with 3D bounding boxes for 23 object classes and accompanied by high-definition map information. The dataset is particularly challenging due to its complex urban environments, featuring dense traffic, frequent pedestrian interactions, and a wide variety of road topologies. Following standard practice, we use a history of 2 s to forecast a future motion of 6 s. We utilize the official dataset split, which comprises 700 scenes for training, 150 for validation, and 150 for testing.

    \item \textbf{NGSIM}: The Next Generation Simulation (NGSIM) dataset~\cite{deo2018convolutional} contains detailed vehicle motion data collected from real-world US highways (US-101 and I-80). It is a widely adopted benchmark for studying vehicle interactions in dense, naturalistic traffic flow. The dataset is renowned for its high frequency of complex maneuvers such as lane changes, merges, and cut-ins. We follow the standard data processing pipeline, where the raw data, recorded at 10 Hz, is downsampled to 5 Hz. For our experiments, we use a history of 3 s (15 frames) to forecast a future of 5 s (25 frames). The dataset is split into training, validation, and testing sets chronologically with a 70\%/10\%/20\% ratio.

    \item \textbf{HighD}: The HighD dataset~\cite{krajewski2018highd} is a large-scale dataset of naturalistic vehicle motions recorded from six different locations on German highways. It features high-quality data captured by drones, resulting in fewer occlusions and measurement errors compared to older datasets. The dataset contains over 110,000 vehicle motions. Similar to our setup for NGSIM, we downsample the data to 5 Hz and use a history of 3 s to forecast the subsequent 5 s. The data is also split into training, validation, and testing sets.
\end{itemize}

\begin{figure*}[htbp]
  \centering
  \includegraphics[width=1\linewidth]{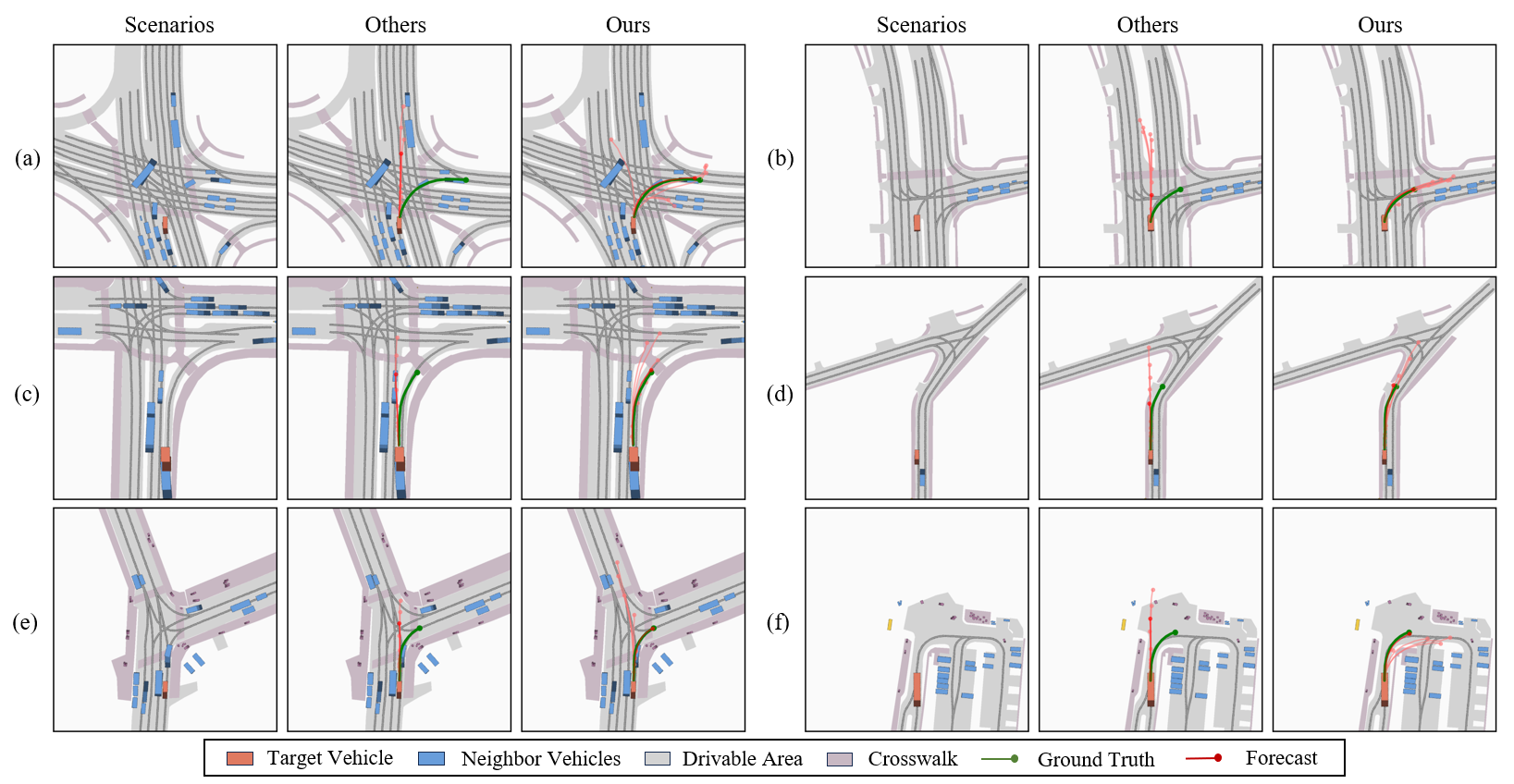}
   \caption{Visualization of multimodal motion forecasting for six right-turn scenarios (a-f) in long-tail urban environments on the nuScenes dataset. Red denotes the highest-probability forecast, and pink represents other multimodal options.}
   \label{Appfig1}
\end{figure*}

\subsection{E.2 Metrics}
To provide a comprehensive and fair assessment of our model's performance, we adopt the standard evaluation metrics widely used for each specific dataset. The primary metrics are detailed below.

\begin{itemize}
    \item \textbf{Minimum Average Displacement Error (minADE$_k$)}: This metric is the standard for evaluating multi-modal forecasts on the nuScenes benchmark. It measures the average L2 distance between the forecasted motion and the ground truth over the entire forecasting horizon. The "minimum" is taken over the $K$ forecasted modes, thus evaluating the performance of the best possible forecast among the $K$ proposals. For a ground truth motion $Y_{gt} = (\mathbf{y}_1, \dots, \mathbf{y}_{T_f})$ and $K$ forecasted motions $\hat{Y}_k = (\hat{\mathbf{y}}_{k,1}, \dots, \hat{\mathbf{y}}_{k,T_f})$, it is defined as:
    \begin{equation}
        \text{minADE}_k = \min_{k \in \{1,\dots,K\}} \left( \frac{1}{T_f} \sum_{t=1}^{T_f} \| \hat{\mathbf{y}}_{k,t} - \mathbf{y}_t \|_2 \right)
    \end{equation}

    \item \textbf{Minimum Final Displacement Error (minFDE$_k$)}: Also a primary metric for the nuScenes benchmark, minFDE$_k$ specifically focuses on the L2 distance at the final time step, $T_f$. It assesses the model's ability to forecast the correct endpoint among its $K$ forecasts. It is defined as:
    \begin{equation}
        \text{minFDE}_k = \min_{k \in \{1,\dots,K\}} \| \hat{\mathbf{y}}_{k,T_f} - \mathbf{y}_{T_f} \|_2
    \end{equation}

    \item \textbf{Miss Rate (MR$_k$)}: A metric for the nuScenes benchmark that measures the percentage of scenes where the final displacement error of all $K$ forecasted modes exceeds a predefined threshold (2 meters). It is defined as:
    \begin{equation}
    \small
        \text{MR}_k = \frac{1}{N} \sum_{i=1}^{N} \mathbb{I}\left( \min_{k \in \{1,\dots,K\}} \| \hat{\mathbf{y}}_{k,T_f}^{(i)} - \mathbf{y}_{T_f}^{(i)} \|_2 > \text{threshold} \right)
    \end{equation}
    where $N$ is the total number of test samples and $\mathbb{I}(\cdot)$ is the indicator function.

    \item \textbf{Root Mean Squared Error (RMSE)}: This metric is the standard evaluation protocol for the NGSIM and HighD datasets. It is used to evaluate the accuracy of the single most likely forecasted motion. RMSE calculates the square root of the average of the squared L2 distances between the forecasted motion $\hat{Y}$ and the ground truth $Y_{gt}$ over all time steps:
    \begin{equation}
        \text{RMSE} = \sqrt{\frac{1}{T_f} \sum_{t=1}^{T_f} \| \hat{\mathbf{y}}_t - \mathbf{y}_t \|_2^2}
    \end{equation}
\end{itemize}

\subsection{E.3 Implementation Details}
Our framework is implemented in PyTorch and trained on a single NVIDIA RTX 3090 GPU. The model's embedding dimension is set to 64. For the Interaction-Aware Encoder, the temporal Transformer consists of 2 layers with 4 attention heads, and the Bayesian MLPs within the Tail Perceiver each have two hidden layers with 128 units. We use the Adam optimizer for all experiments. The main learning rate for the overall model is set to $5 \times 10^{-4}$, while the inner loop learning rate for the MAML-based component is $1 \times 10^{-3}$. For the primary nuScenes benchmark, we train the model for 120 epochs with a batch size of 32. For the NGSIM and HighD datasets, training is conducted for 20 epochs with a larger batch size of 256. Across all datasets, the loss balancing hyperparameters are set to $\lambda_{1}=1.0$ and $\lambda_{2}=1.0$.

\begin{figure*}[htbp]
  \centering
  \includegraphics[width=1\linewidth]{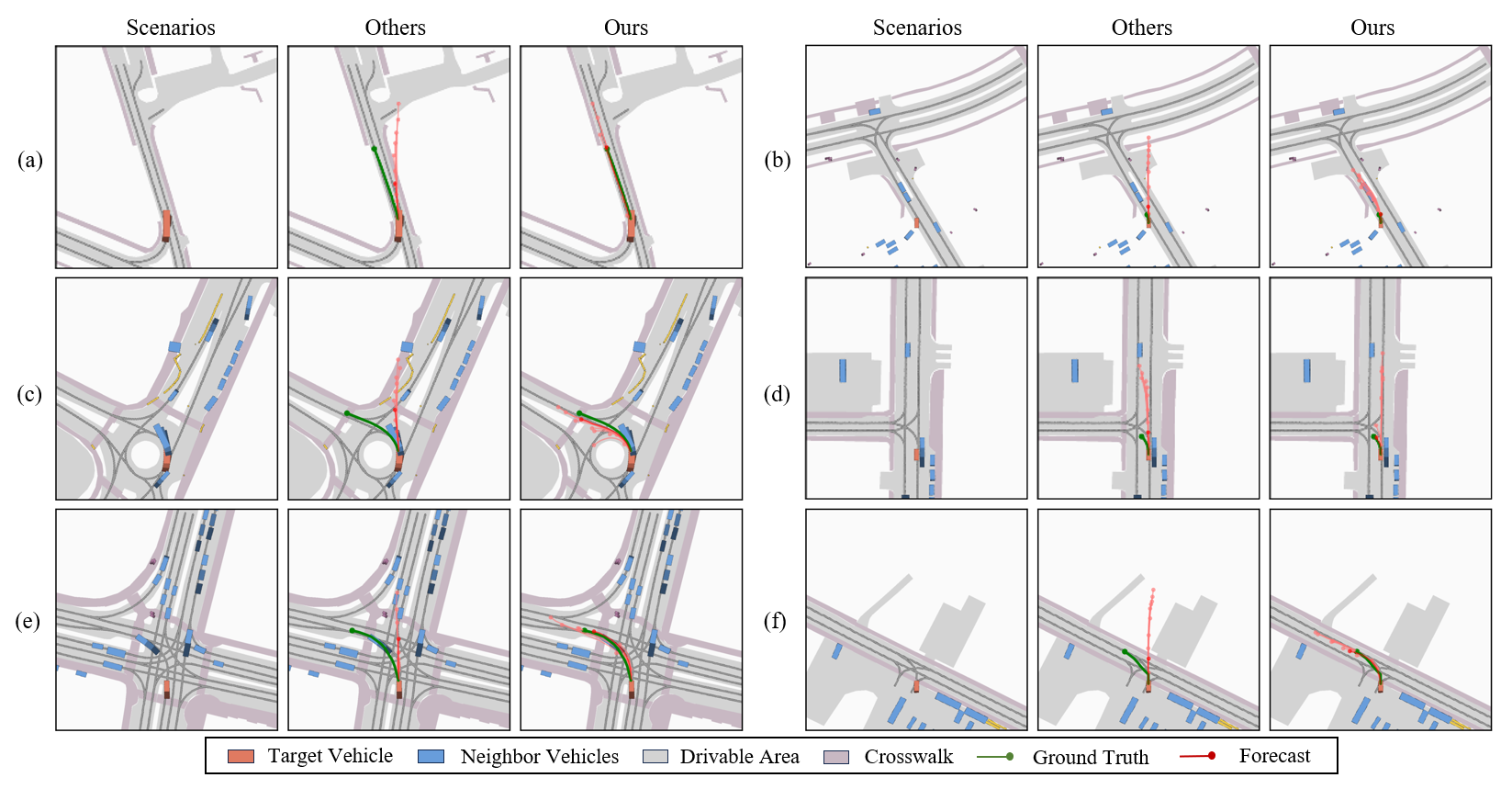}
   \caption{Visualization of multimodal motion forecasting for six left-turn scenarios (a-f) in long-tail urban environments on the nuScenes dataset. Red denotes the highest-probability forecast, and pink represents other multimodal options.}
   \label{Appfig2}
\end{figure*}

\begin{figure*}[htbp]
  \centering
  \includegraphics[width=1\linewidth]{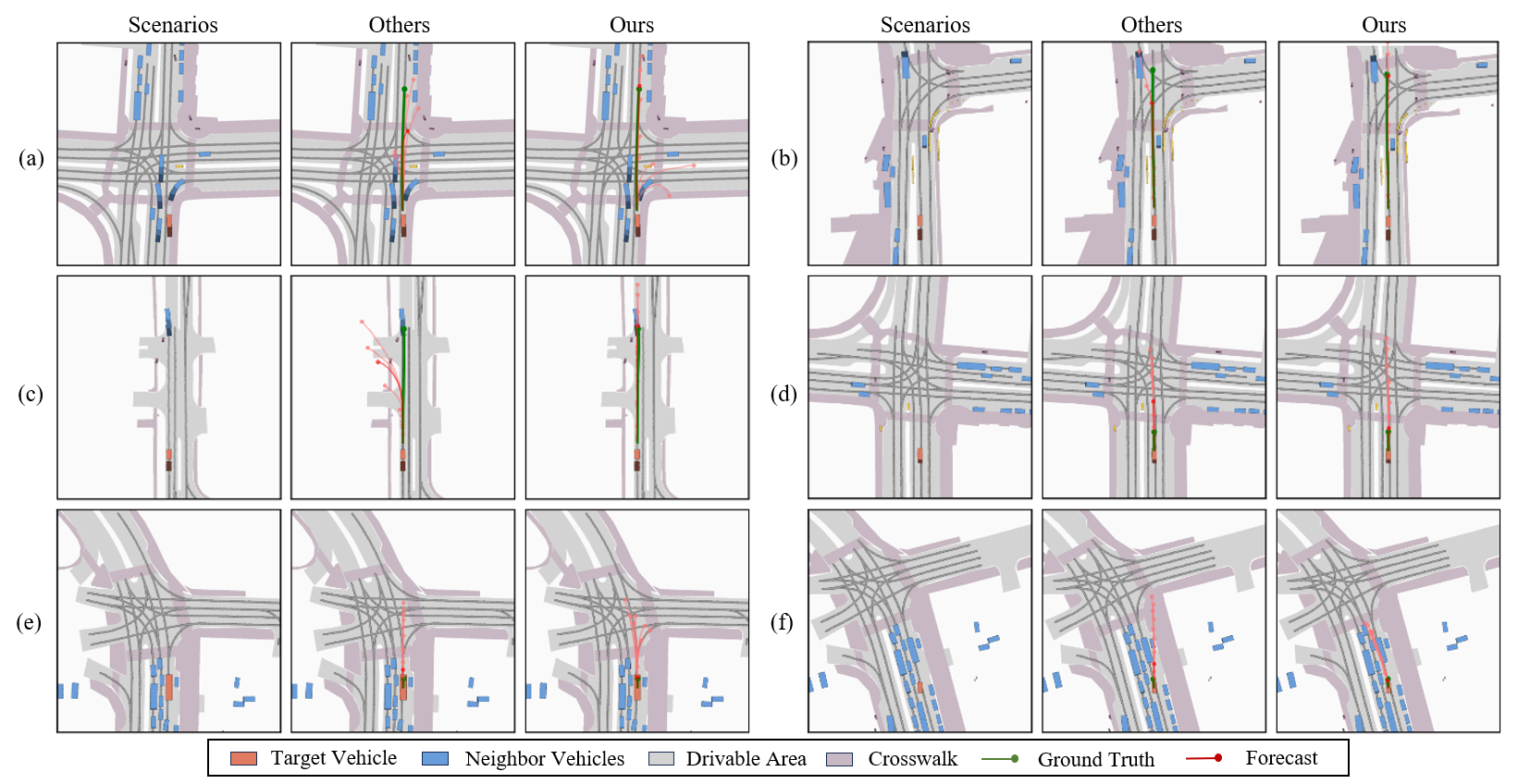}
   \caption{Visualization of multimodal motion forecasting for acceleration (a-c) and deceleration (d-f) scenarios on multi-lane roads on the nuScenes dataset. Red denotes the highest-probability forecast, and pink represents other multimodal options.}
   \label{Appfig3}
\end{figure*}

\section{F. Additional Qualitative Results}

In the Appendix, we provide additional visualizations to further illustrate the superiority of our SAML model in handling long-tail motion forecasting scenarios on the nuScenes dataset. These figures compare SAML with another model \cite{chen2024q}, where the red mode denotes the highest-probability forecast and pink modes represent other multimodal options. Below, we analyze representative cases for right-turn, left-turn, and acceleration/deceleration scenarios.

\subsubsection{Right-Turn Scenarios}
Figure~\ref{Appfig1} presents six right-turn scenarios (a-f) in dense urban environments. The other model struggles with abrupt directional changes, producing clustered pink modes that fail to account for turning behaviors, leading to higher forecasting errors and insufficient coverage of potential turn variations. In contrast, SAML's forecasts align closely with ground truth by leveraging its meta-learning framework to adapt to rare curvatures, generating diverse pink modes that capture a broader range of turn sharpness and contextual interactions, thereby enhancing overall robustness in complex intersections.

\subsubsection{Left-Turn Scenarios}
Figure~\ref{Appfig2} depicts six left-turn cases (a-f). The other model often deviates in its highest-probability mode due to over-reliance on straight-line patterns, producing ordinary straight forecasts that fail to capture turns and limited multimodal coverage, resulting in increased risk in safety-critical situations. SAML excels by incorporating geometric complexity through its Bayesian Tail Perceiver, which improves uncertainty modeling for sparse events and yields more accurate highest-probability modes with realistic alternatives, demonstrating better generalization across varying traffic densities.

\subsubsection{Acceleration/Deceleration Scenarios}
Figure~\ref{Appfig3} shows acceleration in (a-c) and deceleration in (d-f) on multi-lane roads. The other model overestimates or underestimates speed variations, causing forecast overshoots or undershoots from cognitive fixation on common speeds, leading to inaccurate multimodal options in dynamic environments. SAML mitigates this via its Cognitive Set Mechanism, enabling precise capture of dynamic changes and producing multimodal forecasts that better match real-world behaviors, thus improving generalization in long-tail speed-related scenarios with enhanced sensitivity to subtle velocity shifts.

\end{document}